# Intermolecular bond stability of $C_{60}$ dimers and 2D pressure-polymerized $C_{60}$

P. Nagel[1], V. Pasler[1], S. Lebedkin[1], C. Meingast[1], B. Sundqvist[2], T. Tanaka[3], K. Komatsu[3]

[1] *Forschungszentrum Karlsruhe - Technik und Umwelt, Institut für Nukleare Festkörperphysik, PO Box 3640, 76021 Karlsruhe, Germany.*
[2] *Department of Experimental Physics, Umeå University, S-90187 Umeå, Sweden.*
[3] *Institute for Chemical Research, Kyoto University, Uji, Kyoto 611, Japan.*

The thermal stability of $C_{60}$ dimers and 2D pressure-polymerized $C_{60}$ is studied using high-resolution capacitance dilatometry. The transformation of both the dimer and the polymer phases back to 'normal' $C_{60}$ is excellently described by a simple thermally activated process, with activation energies of $1.75 \pm 0.1$ eV (dimer) and $1.9 \pm 0.2$ eV (polymer). These results are compared to previous data of 1D-polymerized $C_{60}$ and photo-polymerized $C_{60}$. The thermal expansivity of the 2D-polymer phase is as much as a factor of ten smaller than that of pure $C_{60}$ and approaches values for diamond.

## INTRODUCTION

Besides the 'normal', face-centered-cubic (fcc), plastic-crystal phase of $C_{60}$, several other phases have been reported in which the $C_{60}$ molecules are linked via covalent bonds to form dimers, linear chains (1D), 2- and even 3-dimensional networks [1]. These polymerized phases can be produced by different methods, such as photo-polymerization (dimers and 1D) [2], high-pressure-temperature conditions (1D, 2D and 3D) [3,4] and the recently published solid-state mechano-chemical reaction of $C_{60}$ with KCN (dimers) [5]. The resulting polymer phases are metastable under ambient conditions and revert back to fcc $C_{60}$ during heating to higher temperatures.

Here we study the kinetics of the polymer to fcc transformation of $C_{60}$ dimers and 2D pressure-polymerized $C_{60}$ using high-resolution dilatometry up to 500 K.

## EXPERIMENTAL

The $C_{60}$ dimers were synthesized by a solid-state mechano-chemical reaction of $C_{60}$ with potassium cyanide [5]. The 2D-polymerized sample was prepared by annealing high-purity sublimed polycrystalline $C_{60}$ at 830 K for 5 hours at a pressure of 2.0 GPa and then cooling the sample before releasing the pressure. Two high-resolution capacitance dilatometers with temperature ranges of 4-300 K [6] and 150-500 K [7], respectively, were used to measure the thermal expansion. Data were taken at constant heating (cooling) rates, and He exchange gas (10 mbar) was used to thermally couple the samples to the dilatometers. To characterize the dimer phase, Raman-spectra were taken with a FT-Raman spectrometer using a Nd:YAG-laser with a wavelength of 1064 nm.

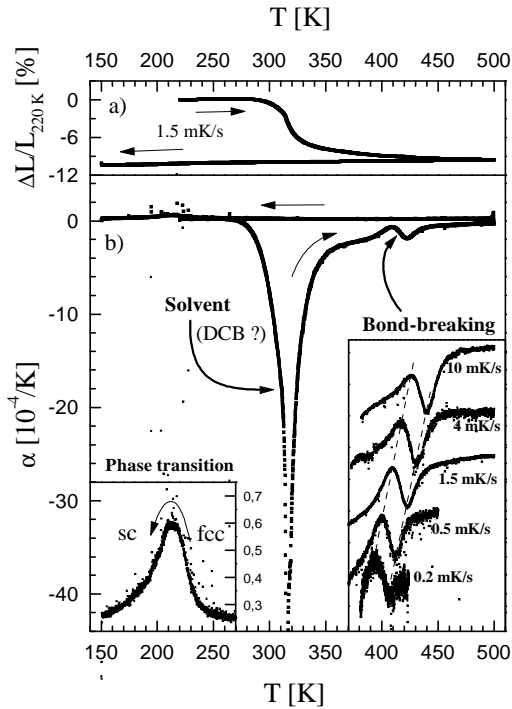

**Fig. 1.** Linear thermal expansion a) and expansivity b) of the $C_{60}$ dimer.

## RESULTS AND DISCUSSION

**$C_{60}$-dimer:** In Fig. 1 we present the linear thermal expansion, $\Delta L/L_0$, and the expansivity, $\alpha = 1/L \cdot dL/dT$, of the $C_{60}$ dimer powder for the first heating cycle from 220 K to 500 K and the subsequent cooling cycle from 500 K to 150 K. Anomalies in $\alpha(T)$ are observed at about 315 K and 415 K in the heating curve and are absent in the cooling curve, indicating irreversible processes. The fcc-sc phase transition near 250 K is recovered in the cooling curve, demonstrating that the dimers have been broken apart by heating to 500 K. In order to determine which of the anomalies upon heating is due to bond breaking, Raman spectra [8] of fresh samples were taken before heating, after annealing at 385 K for three hours and after slowly heating to 500 K. These spectra clearly demonstrate that the 315 K peak is due to evaporating solvent (o-dichlorobenzene) and that bond breaking occurs around 415 K, in good agreement with differential scanning calorimetry results from Ref. 5.

The right inset of Fig. 1b shows that the bond-breaking anomaly shifts to lower temperatures with decreasing heating rates. Defining the bond-breaking temperatures $T_{bb} \equiv T_{bond-breaking}$ as either the maxima or minima of the anomalies allows one to plot the logarithm of the heating rate r versus $1/T_{bb}$, from which an activation energy is obtained. The activation energies are $E_a = 1.70 \pm 0.05$ eV (maxima) and $E_a = 1.79$ eV $\pm 0.05$ eV (minima). To convert the heating rate to a bond breaking rate $\tau_{bb}^{-1}$, the bond breaking was simulated with a simple model in which the fraction of dimers y obeys the differential equation

$$\frac{dy}{dT} = \frac{y}{\tau_{bb}(T) \cdot r} \qquad (1)$$

with an activated rate $\tau_{bb}^{-1}(T) = \nu_0 \cdot \exp(-E_a/k_B T)$. The only free parameter in this model is the attempt frequency, which could be determined to $\nu_0 = (2.6 \pm 3) \cdot 10^{17}$ Hz.

The resulting temperature dependent bond breaking rate of the dimer samples is in very good agreement with that of the photo-polymerized $C_{60}$ films from Y. Wang et al. [9], which strongly suggests that these films consist primarily of dimers. (see Fig. 3)

**2D pressure-polymerized $C_{60}$:** The thermal expansivity of the 2D-polymerized sample measured along two orthogonal directions is shown in Fig. 2. A surprisingly large anisotropy is observed, which possibly could arise from an uniaxial pressure component during synthesis. The expansivity in both directions is significantly smaller than that of both 'normal' $C_{60}$ [10] and 1D-polymerized $C_{60}$ [11], as expected due to stronger covalent bonding between molecules. The values along the direction with the smaller expansivity approach the very small thermal expansivity of diamond [12].

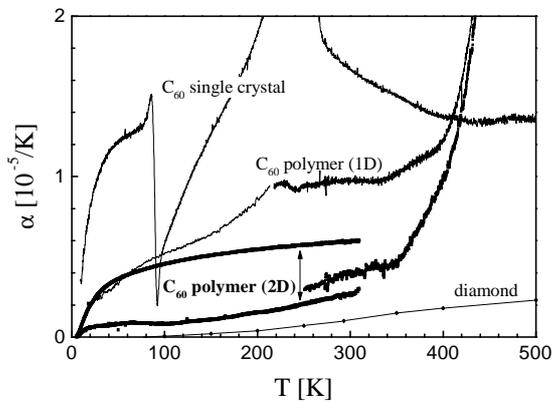

**Fig. 2.** Thermal expansivity of 2D-polymerized $C_{60}$ compared to 'normal' $C_{60}$ [10], 1D-polymerized $C_{60}$ [11] and diamond [12].

The kinetics of depolymerization and the associated volume increase of the 2D-polymerized $C_{60}$ sample was studied in detail by measuring the thermal expansion for repeated heating and cooling cycles between 150 K and 500 K (see Fig. 2 for the initial part of the first heating cycle). We obtained qualitatively very similar results as in our previous measurements on 1D-polymerized $C_{60}$ [11], but with two important differences. First, the length increase upon depolymerization of the 2D polymer was almost a factor of ten larger than for the 1D-polymer (6.3 % vs. 0.7 %), implying a quite large volume increase in good agreement with the literature [3]. Second, although the bond-breaking rate $\tau_{bb}^{-1}$ exhibits an activated behavior with an activation energy $E_a = 1.9 \pm 0.2$ eV for the different heating and cooling cycles just as in the 1D-material, the attempt frequency decreases during the course of the measurement from an initial value close to the value of the 1D-polymer ($\nu_0 = 7 \cdot 10^{15}$ Hz) and ends up with $\nu_0 = 7.3 \cdot 10^{14}$ Hz (see Fig. 3). This may indicate that the sample is a mixture of 1D- and 2D-polymer phases, which are decomposing at different rates. A direct measurement of the time-dependent length-increase at 500 K near the end of depolymerization supports this interpretation; two different relaxation rates (see open-circles in Fig. 3), the values of which excellently match those of the 1D- and 2D-polymers obtained during heating and cooling cycles, are found.

**Comparison:** The bond-breaking rates of dimers, 1D- and 2D-polymerized materials are compared in Fig. 3. Interestingly, $\tau_{bb}^{-1}$ of the dimers is more than 3 orders of magnitude faster than that of the 1D or 2D material. This is quite surprising since experimental results [13] indicate that the dimers are energetically more stable than the 1D or 2D materials. This difference appears to result both from different attempt frequencies and slightly different values of the activation energy. We note that the values of the activation energies found here agree very well with theoretically determined values (~ 1.7 eV) for the 2D-polymer phase [14].

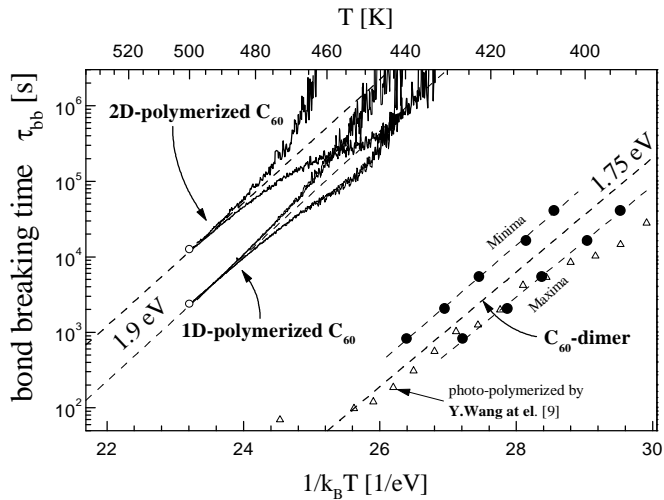

**Fig. 3.** Bond-breaking time $\tau_{bb}$ of the $C_{60}$ dimer (●) and the 2D-polymerized $C_{60}$ compared to 1D-polymerized [11] and photo-polymerized $C_{60}$ (△) [9]. (see text for details)

## CONCLUSIONS

The kinetics of bond-breaking of $C_{60}$ dimers and 2D-polymerized $C_{60}$ has been studied using high-resolution capacitance dilatometry up to 500 K. The bond-breaking rates are thermally activated with activation energies of 1.7 - 1.9 eV. The expansivity of the 2D-polymerized sample is much smaller than the expansivity of both 'normal' $C_{60}$ [10] and even 1D-polymerized $C_{60}$ [11], and approaches values of diamond [12].